\documentclass[prb,twocolumn,showpacs]{revtex4}

\usepackage{amsmath}
\usepackage{graphicx}

\begin{document}

\title{Anomalous Quantum Diffusion at the Superfluid-Insulator
Transition}

\author{Claudio Chamon$^1$ and Chetan Nayak$^2$}

\affiliation{$^1$ Physics Department,
Boston University, Boston, MA 02215\\
$^2$Department of Physics and Astronomy,
University of California at Los Angeles, Los Angeles, CA 90095-1547}

\date{\today }

\begin{abstract}
We consider the problem of the superconductor-insulator
transition in the presence of disorder, assuming that
the fermionic degrees of freedom can be ignored so
that the problem reduces to one of Cooper pair localization.
Weak disorder drives the critical behavior away from the pure
critical point, initially towards a diffusive fixed point.
We consider the effects of Coulomb interactions
and quantum interference at this diffusive
fixed point. Coulomb interactions enhance the conductivity,
in contrast to the situation for fermions, essentially
because the exchange interaction is opposite in sign.
The interaction-driven enhancement of the conductivity
is larger than the weak-localization suppression,
so the system scales to a perfect conductor.
Thus, it is a consistent possibility for
the critical resistivity at the superconductor-insulator
transition to be zero, but this value
is only approached logarithmically. We determine
the values of the critical exponents $\eta,z,\nu$
and comment on possible implications
for the interpretation of experiments.
\end{abstract}


\maketitle


\section{Introduction.}

In a perfectly clean system at $T=0$, the free Fermi gas is perched
precariously at a critical point. An arbitrarily weak interaction will drive
the system superconducting (by the Kohn-Luttinger effect if the interaction
is repulsive). In the presence of disorder, however, the {\em diffusive}
Fermi liquid is a stable phase for a finite range of interaction and disorder
strengths in dimensions $d>2$. In $d=2$, it remains an open problem whether
or not fermions have a stable diffusive metallic phase.  Such a phase, if it
exists, could not be adiabatically connected to the Fermi liquid
\cite{Dobrasavlevic97} since the non-interacting Fermi gas is always
insulating in the presence of disorder in $d=2$. In the limit of weak
disorder, this can be understood as a quantum interference effect which is
singular as a result of the diffusive nature of electron propagation in a
disordered system: diffusion at intermediate length scales (longer than the
elastic mean-free path) thwarts diffusion at long scales (longer than the
localization length) \cite{Abrahams79}. The interacting-electron problem
remains unresolved because interactions in the spin-triplet channel are also
singular as a result of the languid pace of diffusive
motion\cite{Finkelstein83,Castellani84,Belitz94,Chamon99}.  The upshot of the
interplay between these different singularities is unknown (see, however,
ref. \onlinecite{Chamon00,Kirkpatrick96}).

Consider the critical point separating the insulating
and superfluid phases of a perfectly
clean system of bosons at $T=0$ in $2D$. We would like to
draw an analogy between it and the free Fermi gas.
In the bosonic case, there is a particular
value of the chemical potential
for which the system has gapless critical modes, loosely
analogous to the excitations of the free Fermi gas.
For any other value of the chemical potential,
the bosons are either in a superfluid state
 -- a superconducting state, if we assume that the bosons
are Cooper pairs -- or in a gapped insulating state. Suppose we now
add disorder to this system. What is the fate
of this critical point? On general grounds, we believe
that it is unlikely to broaden into a stable diffusive metallic phase,
and that the only stable phases are insulating
(Mott insulator or Bose glass \cite{Fisher89b}) or superconducting.
Instead, we expect a diffusive metallic
critical point with a {\em universal} conductivity
separating the insulating and
superconducting phases.
The analogy between Fermi and Bose systems
is imprecise, but it emphasizes the important
point that in both cases there is a ballistic critical point
in the clean system which must be usurped by a
diffusive fixed point in the disordered one.

Such a fixed point should be amenable to
analysis by methods similar to those used for
the diffusive Fermi liquid. Conversely,
expansion about the pure critical point -- which
is ballistic, not diffusive -- should fail.
In considering such a perspective, one is faced with
the following question: why do quantum interference effects,
which appear to be such an inevitable consequence of
diffusive motion, not preclude a finite conductivity
at the superfluid-insulator transition?
The answer must lie in the effects of interactions,
which one might hope to tame since spinless bosons,
such as singlet Cooper pairs, do not have a
triplet channel -- the troublesome,
singular one -- through which to interact.

In this paper, we present the results of such
an analysis. We find that there are two competing
effects at a putative $2D$ diffusive Bose liquid
critical point: one resulting from interactions between the
bosons; the other, from quantum interference,
i.e. weak localization. In the fermionic case, it is advisable to
consider quantum interference and interactions
on the same footing since they lead to similar
logarithmic corrections at the perturbative
level. In the bosonic case, one must perforce
do so, since quantum interference leads to the existence
of localized states even in the weak disorder limit,
and bosons would congregate in the
lowest energy localized state in the absence of interactions.
We find that the effect of interactions is stronger
than quantum interference and drives the system to a perfect conductor,
thereby explaining how diffusion can remain
impervious to localization. This result is congenial
to one's intuition that repulsive interactions
should disfavor localization. Potential wells due to
impurities diminish in attractiveness when they
are occupied and, as a result, the random potential is effectively screened.
This effect is present for both short-ranged interactions
as well as long-ranged Coulomb interactions, but is stronger
in the latter case. The same phenomenon occurs in fermionic systems
as well, but it competes with the exchange part of the interaction,
which is opposite
in sign due to Fermi statistics. If the interaction is short-ranged,
it is irrelevant for spinless fermions, so it has
no effect on the conductivity in the infrared limit. (This
is clear in the $\delta$-function limit, where
the direct and exchange interactions cancel.)
In the case of Coulomb interactions, the exchange interaction between
spinless fermions dominates and suppresses the conductivity.
In the case of spin-$1/2$ fermions, the runaway flow of
the triplet interaction amplitude
indicates that the Hartree interaction
begins to prevail over the exchange interaction at longer length
scales, thereby leading to an enhanced conductivity. However,
the interaction strength diverges before a metallic fixed
point is reached, and no conclusion can be drawn about the existence of a
metallic state at zero-temperature. These
difficulties do not arise in the bosonic case. The exchange
interaction has the same sign as the direct one, and both
enhance the conductivity.

Our result is valid for large conductivities in units
of ${e^2}/h$. Hence, if the bare conductivity is large -- as it can be if
the bosons have an anisotropic mass tensor --
then the renormalized conductivity is infinite.
If the bare conductivity is small, then there are two
possibilities. If the conductivity initially flows to sufficiently large
values that we can apply our calculation, then it will
continue to flow to infinity. However, it is also possible that
the system will flow in this case to
a different fixed point at which the
conductivity if finite. In such a scenario, there would be two
different possible universality classes of superconductor-insulator
transitions. In either case, we conclude that it is a consistent
possibility for the critical point between the
superfluid and insulating states of
a disordered Bose liquid to be a {\em perfect conductor}.

We derive these results in a non-linear $\sigma$-model (NL$\sigma$M)
formulation of the problem of diffusing, interacting bosons.
Our NL$\sigma$M is very similar to Finkelstein's model
for fermions \cite{Finkelstein83}. However, the NL$\sigma$M plays a very
different role in this problem than in the fermionic problem.
There, the NL$\sigma$M describes the entire metallic phase.
In $2+\epsilon$ dimensions, the metal-insulator transition
occurs near the metallic fixed point, so the NL$\sigma$M
excompasses it as well. In the bosonic problem
which models the superconductor-insulator transition,
our NL$\sigma$M describes the {\it critical point}. The
antiferromagnetic Heisenberg model in
$d>2$ provides an enlightening analogy.
For isotropic exchange coupling ${J_z}={J_{x,y}}$,
the model is ordered and is described by a NL$\sigma$M.
In the ordered phase, continuous symmetries are broken
so there are Goldstone modes; this is the analog of our
critical point. For ${J_z}>{J_{x,y}}$, the model
develops Ising order with a gap; this is analogous to our
insulating phase. For ${J_z}<{J_{x,y}}$, the model
develops $XY$ order, which is analogous
to our superconducting phase.

\section{Dirty Bosons}

Following \onlinecite{Fisher89b}, we will treat the Cooper
pairs in a dirty superconductor as bosons moving in a random
potential. We will assume that all fermionic degrees of
freedom are gapped or localized and are therefore unimportant.
This assumption has been called into question recently
\cite{Valles,Kapitulnik01}. If fermionic degrees of freedom prove
to play an important role at the superconductor-insulator transition,
then our analysis will need to be modified to include them,
but our description of dirty bosons will remain an
important component of a richer
description of the superconductor-insulator transition.

Note that we are studying here the generic transition \cite{Fisher89b}
between the Bose Glass and superfluid phases which occurs
at an incommmenusurate boson density. In the special case
in which there are an integer number of bosons per lattice site,
there may be a direct transition between Mott Insulating
and superfluid phases which is tuned by varying the ratio of
the hopping and interaction parameters \cite{Kisker97}.

We begin with a system of interacting bosons moving
in a random potential in two dimensions. The derivation
which follows goes through in arbitrary dimension
with minor changes, but $d=2$ is the most interesting
case. The imaginary-time action is:
\begin{multline}
S = \int {d^2}x\,d\tau\:{\psi^*}\left({\partial_\tau}-
\frac{1}{2m}{\nabla^2}-\mu + V(x)\right)\psi\\
+ \int {d^2}x\,{d^d}x'\,d\tau\:
{\psi^*}(x)\psi(x) u(x-x'){\psi^*}(x')\psi(x')
\end{multline}
$u(x-x')$ is the interaction between bosons; we will consider
the cases of both short-ranged interactions and Coulomb interactions.
$V(x)$ is the random potential; we use the replica trick to average over
it, thereby obtaining the action:
\begin{multline}
S = \int {d^2}x\,d\tau\:{\psi_a^*}(x,\tau)\left({\partial_\tau}-
\frac{1}{2m}{\nabla^2}-\mu\right){\psi_a}(x,\tau)\\
- \int {d^2}x\,d\tau\,d\tau'\: \frac{1}{2} v_0\:
{\psi_a^*}(x,\tau){\psi_a}(x,\tau)
{\psi_b^*}(x,\tau){\psi_b}(x,\tau')\\
+ \int {d^2}x\,{d^2}x'\,d\tau\:
{\psi_a^*}(x){\psi_a(x)} u(x-x'){\psi_a^*}(x'){\psi_a}(x')
\end{multline}
$a=1,2,\ldots,N$ is a replica index.
We have assumed that the potential has the Gaussian white-noise
distribution $\overline{V(x)V(x')}=v_0\,\delta(x-x')$. 

This action is problematic because it is not positive definite
as a result of the second term. To cure this, we will rotate the
integration contour in the functional integral, as one does in the
non-interacting case. This can be done more conveniently if we work in
the Matsubara frequency representation and separate the real
and imaginary parts of the Matsubara fields
$\psi_{na}=\phi_{na1}+i\phi_{na2}$, where ${\epsilon_n}=2\pi n/\beta$.
The action can be made positive definite by rotating the fields
in the following way: ${\phi_{naA}}\rightarrow
{e^{-{\frac{\pi}{4} i\, \text{sgn}(n)}}{\phi_{naA}}}$, $A=1,2$. We
rotate the $n=0$ mode along with the $n>0$ modes.
\begin{widetext}
The action now takes the form:
\begin{multline}
\label{eq:boson-action}
S = {\sum_{n,m}}\int {d^2}x\:i{\phi_{naA}}
(x,\tau)\left(i{\epsilon_n}+
\frac{1}{2m}{\nabla^2}+\mu\right)\,
{\Lambda_{nm}}{\phi_{maA}}(x,\tau)\\
+ \sum_{n,n',m,m'}\int {d^2}x\: \frac{1}{2}v_0\:
{\phi_{naA}}{\Lambda_{nn'}}{\phi_{n'aA}}
{\phi_{mbB}}{\Lambda_{mm'}}{\phi_{m'bB}}\\
+ \sum_{n_1,\dots,n_4}
\int {d^2}x\,{d^2}x'\:\left[{e^{-\pi i \sum\text{sgn}({m_i})/4}}\right]
{\phi_{{m_1}aA}}(x){\phi_{{m_1}aA}(x)}\,
 u(x-x') {\phi_{{m_1}aB}}(x'){\phi_{{m_1}aB}}(x')
\end{multline}
where ${\Lambda_{mm'}}=\text{sgn}(m)\,\delta_{mm'}$.
\end{widetext}

In the absence of disorder, repulsive interactions are marginally
irrelevant, and the critical behavior of (\ref{eq:boson-action})
is controlled by the Gaussian fixed point \cite{Fisher89b}.
Now consider a perturbative treatment of the disorder.
In the self-consistent Born approximation, we find
a self-energy due to disorder of the form:
\begin{multline}
\Sigma\left({\epsilon_n}\right) = 
\frac{mv_0}{2\pi}\Biggl[
\ln\left|\frac{{\Lambda^2}/2m}{i{\epsilon_n}+
\mu+\Sigma\left({\epsilon_n}\right)
}
\right| +\\
i\,\tan^{-1}\left(\frac{{\epsilon_n}+\text{Im}
\Sigma\left({\epsilon_n}\right)}{\mu+\text{Re}
\Sigma\left({\epsilon_n}\right)}\right)
\Biggr]
\end{multline}
The random potential shifts the chemical potential
and also gives the bosons a finite lifetime $\tau$.
As a result of the lifetime $\tau$, single-boson excitations
are no longer long-lived degrees of freedom.
However, particle-hole pairs are long-lived, as may be seen from
the conductivity which, at this level of
approximation, is $\sigma=\frac{1}{2\pi^3}$.

This does {\it not} preclude critical behavior in the single-particle
properties, as has already been seen in the context
of interacting fermions \cite{Finkelstein83} and of
quasiparticles in a disordered $d$-wave superconductor
where there are density-of-states corrections
and also in the context of non-interacting electrons
with an extra sublattice symmetry, where the single-particle
Green function itself is critical \cite{Altland00}.

The conductivity is small because there are no particle-hole pairs for
$\tau=\infty$ (since the transition occurs at the bottom of a quadratic
band). A finite lifetime leads to a small density of states $\sim 1/\tau$ for
particle-hole pairs, which cancels the factor of the lifetime to which
$\sigma$ is customarily proportional, thereby leading to a conductivity which
is $O(1)$. However, we note that a parametrically large conductivity can be
obtained in a slight generalization to a model of two species of bosons with
anisotropic masses and that mix upon scattering. Suppose that one of them has
${m_x}={m_1}$, ${m_y}={m_2}$, while the other has masses reversed. Then we
find that $\sigma=\left[\sqrt{{m_1}/{m_2}} +
  \sqrt{{m_2}/{m_1}}\right]/2{\pi^3}$.  For sufficiently large or small ratio
${m_1}/{m_2}$, the conductivity will be large. Such a situation could occur,
for instance, in a two-band model in which the two bands of electrons have
anisotropic masses, leading to anisotropic masses for the Cooper pairs.

An RG analysis of the dirty boson problem
yields the following RG equation
in an $\epsilon$-expansion about $d=4$
\cite{Fisher89b}:
\begin{equation}
\frac{dv_0}{d\ell} = \left(\epsilon+{\epsilon_\tau}\right)v_0
+ B{v_0^2} + \ldots
\end{equation}
with $4-\epsilon-{\epsilon_\tau}$
spatial dimensions and ${\epsilon_\tau}$
time dimensions (the interesting case $d=2$ occurs
at $\epsilon={\epsilon_\tau}=1$). $B>0$, so there
is no fixed point at weak coupling; instead, there
is a runaway flow to strong disorder. We interpret
this as an instability of the pure critical point,
at which the critical modes are ballistic, to the diffusive fixed point.
To access the latter fixed point, we will construct
a non-linear $\sigma$ model which is appropriate for physics at
length scales longer than the mean-free path.
In this regime, transport is diffusive,
and we may neglect degrees of freedom,
such as the $\phi$ fields, which are short-lived.

\section{Saddle-Points for Dirty Bosons}

In the absence of the $i\epsilon_n$ term,
the non-interacting part of the action
(\ref{eq:boson-action}) has an $O\left((k+1)N,kN\right)$ symmetry,
where $k$ is a cutoff on the Matsubara frequencies. The key assumption
of Finkelstein's theory \cite{Finkelstein83}
for fermions is that the elevation
of the energies of the
diffusion modes by the $i\epsilon_n$ term and the
interactions can be neglected
compared to the gaps associated with other degrees of freedom; when this
condition is satisfied, it is valid to retain only interacting diffusion modes
and ignore all other degrees of freedom. We make the
same assumption here in our description of the critical point.
In the superfluid state, this is clearly not sufficient,
and we will have to retain an extra degree of freedom.
It may also be necessary to include extra degrees of
freedom to properly describe the Bose glass insulating
state.

Our treatment of the critical saddle-point and
non-linear $\sigma$-model (NL$\sigma$M)
for interacting bosons follows that of Finkelstein for the
fermionic case and also that of the bosonic representation
of the non-interacting problem. Hence, we will merely give
an outline in this section and the next, emphasizing the important differences.
Details are presented in appendix \ref{appendix-Q-sigma}.

We begin by using the Hubbard-Stratonovich transformation
to decouple the $v_0$ term with a matrix $Q^{mn}_{ab,AB}$.
We then decouple the interaction in two different ways with $X$,
which decouples the direct and exchange channels according
to $X\sim {\psi^*}\psi$, and $X_c$, which decouples the
Cooper channel according to ${X_c}\sim \psi\psi$.
Finally, we decouple the chemical
potential term with $\Phi\sim \psi$.
In this way, we have a
system of non-interacting bosons at zero chemical potential
-- their critical point -- moving in the background fields
$X$, $X_c$, and $\Phi$. Integrating out the $\phi$ fields,
we obtain the effective action (see
appendix \ref{appendix-Q-sigma}):
\begin{widetext}
\begin{eqnarray}
{S_{\text eff}}[Q,Y,Z,Z^\dagger,\Phi] =&&\\
&&\!\!\!\!\!\!\!\!\!\!\!\!\!\!\!\!\!\!\!\!\!\!\!\!\!\!\!\!\!\!\!\!\!\!\!\!
\sum\int\biggl[\text{tr}\,\ln\left(i{\epsilon_n}+
\frac{1}{2m}{\nabla^2} + Q 
-i \sqrt{2\Gamma} 
\;{e^{-i\frac{\pi}{4}\Lambda}} \;X\;{e^{+i\frac{\pi}{4}\Lambda}} 
-i \sqrt{2\Gamma_c} 
\;{e^{-i\frac{\pi}{4}\Lambda}} \;\frac{1}{2}({X_c}+{X_c}^\dagger)
\;{e^{+i\frac{\pi}{4}\Lambda}} 
\right)\\
&&+ \frac{1}{2v_0}\text{tr}\left({Q^2}\right) +
\frac{1}{2}\text{tr}\left({X^2}\right) +
\frac{1}{2}\text{tr}\left({{X_c}^\dagger {X_c}}\right)\\
&&+ \mu ({\Phi^*}\:\Phi)\:\hat{G}\: 
\left( \begin{array}{l}{\Phi^*}\\{\Phi}\end{array} \right)
\biggr]
\label{eqn:H-S-action}
\end{eqnarray}
\end{widetext}
The Green function $\hat{G}$ of the $\phi_A$s is written as
a $2\times 2$ matrix in the final line to emphasize the
particle-hole structure. It is the operator inverse of the expression
inside the logarithm. For $\mu\leq 0$, it is not even necessary to
introduce $\Phi$; we can simply drop the last line
of (\ref{eqn:H-S-action}) and insert $\mu$ inside the logarithm.

Let us now consider the saddle-points of this effective action.
For $\mu>0$, there is a saddle-point with
$\left\langle\Phi\right\rangle\neq 0$. (When we include fluctuations,
$\mu$ will be renormalized, so the critical value will not be
zero.) When $\Phi$ develops an expectation
value, $Q$, ${X_c}$, and $X$ are forced to follow since they are
coupled directly to bilinears in $\Phi$. This is the superfluid phase.

For $\mu\leq 0$, let us consider the non-interacting case
$\Gamma={\Gamma_c}=0$. The saddle-point condition is
\begin{equation}
\label{eqn:S-P-condition}
\hat{Q}=-{v_0}\int \frac{{d^2}p}{(2\pi)^2}\,
\frac{1}{i\hat{\epsilon_n}+\frac{1}{2m}{\nabla^2}\, +
\mu + \hat{Q}}
\end{equation}
Let us absorb the real part of the saddle-point value of 
$Q$ into a renormalized $\mu_R$ and focus on the imaginary
part. 

For ${\mu_R}=0$, the saddle-point solution of (\ref{eqn:S-P-condition})
is
\begin{equation}
Q^{m,n}_{ab,AB} = 
i\; \frac{mv_0}{2}\,\text{sgn}\left({\epsilon_n}\right)
\;\delta_{mn}\;\delta_{AB}\;\delta_{ab}
\label{eq:saddle-sol}
\end{equation}
This is the diffusive saddle-point for self-consistent Born scattering of
critical bosons by impurities. It corresponds to a finite density of states
for the bosons at this level of approximation. Notice that this saddle-point
solution is taken to be replica symmetric.

Now, for ${\mu_R}<0$, there is another translationally-invariant saddle-point
with $Q=0$. For this solution, a non-zero density-of-states in not generated
in the insulating state at this level of approximation; it remains a Mott
insulator. We would like to point out two possible mechanisms to generate the
finite density of states that occurs in the Bose-glass phase. One is that the
correct saddle points are replica symmetry broken mixtures of the $Q=0$ and
Eq.~\ref{eq:saddle-sol} solutions. A possible self-consistent solution is one
still diagonal in replicas but with zero matrix elements for $p$ replicas and
unit matrix elements for $n-p$ replicas. Another possibility is that there
are non-trivial instanton saddle-points which generate a finite
density-of-states \cite{Cardy78}. In the absence of interactions, the bosons
will condense into these localized states, so we must consider the
corresponding instantons with $\Gamma,{\Gamma_c}\neq 0$. At present, we do
not have a description of the Bose glass insulator, but this does not affect
our ability to describe the critical point between it and a superconductor.

It is useful, in thinking about this theory, to imagine
lowering the temperature of a system of dirty bosons.
At finite temperature, there will be a finite wedge
in the phase diagram -- the quantum critical region
\cite{Chakravarty89,Sachdev99} -- where the bosons will
be effectively critical. In this regime, we may begin
by considering non-interacting bosons which are semiclassically
scattered by impurities. As we decrease the temperature,
we must begin to include the effects of interactions
and of quantum interference processes. If we stray too far
from the critical $\mu$ as we lower the temperature,
thereby leaving the quantum critical region, then we cannot include
these effects perturbatively. It is clear that they completely
destabilize the diffusive saddle point, so they must be included right from
the start (e.g. by starting from new saddle-points,
as we have sketched above) in order to describe the
superfluid or insulating phases
correctly. However, so long as we remain at criticality,
we can hope to account for these effects perturbatively.
To such an analysis we turn in the next section.

\section{$\sigma$-Model for Interacting Bosons}

To go beyond a non-interacting, semiclassical
analysis and include the
effects of interactions and quantum interference,
we construct the NL$\sigma$M which accounts for fluctuations
of $Q$.
We shift $Q$ by $i{\epsilon_n}
+ \sqrt{2\Gamma} \;{e^{-i\frac{\pi}{4}\Lambda}}
\;X\;{e^{+i\frac{\pi}{4}\Lambda}} 
+ \sqrt{2\Gamma_c} \;{e^{-i\frac{\pi}{4}\Lambda}}
\;\frac{1}{2}({X_c}+{X_c}^\dagger)\;{e^{+i\frac{\pi}{4}\Lambda}}$
to remove these terms from the $\text{tr}\ln[\cdot]$.
Then, we expand the $\text{tr}\ln[\cdot]$ about the saddle point
and integrate out $X$, ${X_c}$. We obtain an effective action
which is essentially the same as Finkelstein's action
for the fermionic problem (see Appendices \ref{appendix-Q-sigma}
and \ref{appendix-diffusion}):
\begin{widetext}
\begin{eqnarray}
\label{eqn:sigma-model}
{S_{\text eff}}[Q] =&& 
\int d^d x\; 
\biggl\{D\,\text{tr}{\left(\nabla Q\right)^2}
- 4iZ \text{tr}\left(\hat{\epsilon}Q\right) \\
&&+\Gamma
\sum_{n_1,\dots,n_4}\;
\left[
{e^{+i\frac{\pi}{4}n_1}}\;
Q^{n_1n_2}_{aa,AA'} \;{e^{-i\frac{\pi}{4}n_2}}
\right]\;
J_{AA'}\;J_{BB'}\;
\left[{e^{+i\frac{\pi}{4}n_3}}\;
Q^{n_3n_4}_{aa,BB'} \;{e^{-i\frac{\pi}{4}n_4}}
\right]\;
\delta_{n_1-n_2+n_3-n_4}
\nonumber\\
&&+\Gamma_{c}
\sum_{n_1,\dots,n_4}\;
\left[
{e^{+i\frac{\pi}{4}n_1}}\;
Q^{n_1n_2}_{aa,AA'} \;{e^{-i\frac{\pi}{4}n_2}}
\right]\;
S^+_{AA'}\;S^-_{BB'}\;
\left[{e^{+i\frac{\pi}{4}n_3}}\;
Q^{n_3n_4}_{aa,BB'} \;{e^{-i\frac{\pi}{4}n_4}}
\right]\;
\delta_{n_1+n_2-n_3-n_4}
\biggl\}
\nonumber
\end{eqnarray}
\end{widetext}
where $J_{AB}=\frac{1}{\sqrt{2}} (\delta_{AB}-\sigma_{AB}^2)$ and
${S^\pm_{AB}}={\sigma^3_{AB}}\pm i{\sigma^1_{AB}}$ express the particle-hole
matrix structure for the density-density and Cooper channels, respectively.
The parameter $Z$ is $1$ in the bare action above; however, this quantity is
renormalized, so we have introduced it explicitly here.  We have absorbed the
density-of-states into the diffusion constant $D$ (and also the coefficients
of the other terms); the resulting quantity is just the bare conductivity and
is given by $D=1/2{\pi^3}$ in the above model.  However, as we noted earlier
by considering a model with anisotropic masses, and a sufficiently large or
small ratio ${m_1}/{m_2}$, the bare conductivity will be large. The
resistivity $g=1/(2\pi D)$ is the expansion parameter used in our RG
equations, so this observation gives us a limit in which they can be applied
without apology.

It may strike the reader as strange that we are using
a NL$\sigma$M to describe a critical point; usually
NL$\sigma$Ms are used to describe stable phases because
they are so highly constrained by symmetry. However,
the NL$\sigma$M of eq. \ref{eqn:sigma-model} is not, in fact,
so rigidly constrained at all.  The interaction
terms and the $\text{tr}\left(\hat{\epsilon}Q\right)$
term explicitly breaks the $O\left((k+1)N,kN\right)$ `symmetry'
of the model. The latter breaks it in such a way as to push
the theory into a diffusive metallic state. However,
this symmetry-breaking `field' is small in the low-energy
limit, so other symmetry-breaking fields (or anisotropies)
can intervene instead. When $\Phi$ orders in eq. \ref{eqn:H-S-action},
$Q$ is forced away from the diffusive `direction' in
its saddle point manifold, and into the superfluid `plane',
where $Q$ has non-vanishing components which are
off-diagonal in particle-hole indices. {\it Thus, we can understand
the perturbations which lower the symmetry of the saddle-point
manifold as perturbations which drive the system away from criticality.}
There are a variety of ways in which one can imagine
driving the system into an insulating phase. In the absence
of a better understanding of the Bose glass phase, we consider
the simplest which is just a `mass' term of the form
$\text{tr}\left(M\;Q\right)$, with $M$ a constant matrix say in replica
space, which breaks the symmetry of
the saddle-point manifold and leads to an insulating state.
Such a perturbation differs only in index strucure with
the one imposed by a finite $\Phi$. Such a term is
also generated by shifting $\mu$ out of the $\text{tr}\ln[.]$
term when considering replica symmetry broken saddles.
Note that none of these possibilities can occur in the
non-interacting problem, where the symmetry of the
saddle-point manifold is a genuine symmetry.

We parametrize $Q$ about the non-interacting saddle point as
\begin{eqnarray}
\label{eqn:Q-param}
{Q} = \frac{mv_0}{2}\,
\left(\begin{array}{lccr}
i\left(1+q{q^T}\right)^{1/2} & q \\
{q^T} &  -i\left(1+{q^T}q\right)^{1/2}\\
                      \end{array} \right)
\end{eqnarray}
where the block structure is in frequency space, {\it i.e.}, the matrix
$q_{nm}$ is such that $n\ge 0$ and $m<0$.

The resulting action is very similar to the $O(N)$ sigma model which is
appropriate for a system of fermions with spin-orbit scattering. Indeed, one
can be transformed into the other by redefining $q\to q$, $q^T\to-q^T$, and
$D\rightarrow -D$.  The interaction terms look somewhat strange at first
glance, but the extra $i$'s in (\ref{eqn:Q-param}) are precisely compensated
by the explicit factors of ${e^{\pm i{\frac{\pi}{4} n_i}}}$ in
Eq.~(\ref{eqn:sigma-model}) (see appendix \ref{appendix-saddle-param}).

\section{RG Equations}

Taking advantage of the observation at the end of the
previous section, we can obtain the RG equations for our $\sigma$-model
by flipping $g\rightarrow -g$ in the equations for the corresponding
fermionic model. Some factors of $2$ will be different because
our bosons are spinless. More details may be found in appendix
\ref{appendix-saddle-param}.

The RG equation for $\Gamma_c^2$ is:
\begin{equation}
\frac{d\Gamma_c}{d\ell} = -g{\Gamma_c}-{\Gamma_c^2}
\end{equation}
Observe that $\Gamma_c$ flows to zero, even if
$g=0$. Hence, we set $\Gamma_c$ to its fixed point value
of zero and consider the RG equations for $g$, $\Gamma$, and $Z$
in its absence. To order $g^2$ and all orders in $\Gamma$
(although, of course, we cannot access non-perturbative
effects associated with saddle-points which are far from
the non-interacting diffusive one), the RG equations are:
\begin{eqnarray}
\frac{dg}{d\ell} &=& \frac{1}{2}\,{g^2} - {g^2}
\left[2+2\left(\frac{Z}{\Gamma}-1\right)\,
\ln\left(1-\frac{\Gamma}{Z}\right)\right]
\label{eqn:g-rg}\\
\frac{dZ}{d\ell} &=& g\,\Gamma
\label{eqn:Z-rg}\\
\frac{d\Gamma}{d\ell} &=& g\,\Gamma
\label{eqn:Gamma-rg}
\end{eqnarray}

\begin{figure}[tbh!]
\includegraphics[width=3.25in]{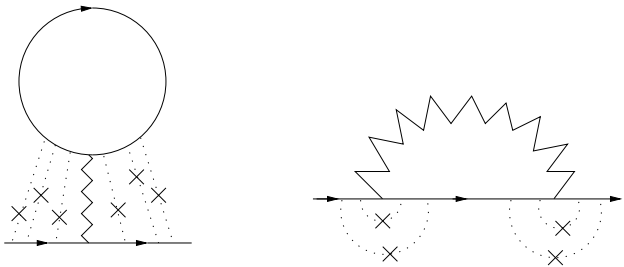}
\caption{The Hartree and Fock diagrams for the boson
self-energy}
\label{fig:ha-fock}
\end{figure}

The physics of these equations is clear from the discussion
in the introduction. Interactions always enhance the conductivity
to order $g^2$ because the exchange term has the same sign as the
direct term (they are folded into a single $\Gamma$ in the
bosonic NL$\sigma$M (\ref{eqn:sigma-model})).
The gist of the effect can be seen from
the Hartree and Fock diagrams for the boson self-energy
displayed in fig. \ref{fig:ha-fock}.
In the Hartree diagram, the boson line is repelled
by the boson bubble which is a measure of the ground
state density. (In a pure system, this is uniform and cancelled
by the neutralizing background.) In a fermionic system,
the Fock diagram comes with the opposite
sign, so it is an effective attraction. In a bosonic
system, however, both diagrams come with the same sign
and lead to a repulsion of particles from regions
of high density -- which, of course, are precisely the
regions where there are deep wells in the random potential.

The interaction strength, $\Gamma$,
grows in importance at low energies
because it plays a role somewhat analogous to the Pauli exclusion
principle: in its absence, all of the bosons would sit in the
lowest minimum of the random potential. $Z$ must follow $\Gamma$
in order to maintain a finite compressibility.

\begin{figure}[tbh!]
\includegraphics[width=3in]{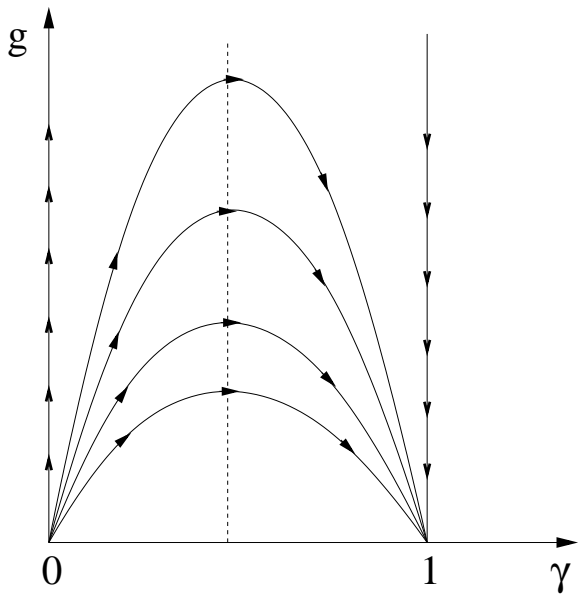}
\caption{RG flow for the resistivity $g$ and interaction 
parameter $\gamma=\Gamma/Z$}
\label{fig:flow}
\end{figure}

Notice from Eqs.~(\ref{eqn:Z-rg},\ref{eqn:Gamma-rg}) that
$Z-\Gamma$ remains invariant under the RG flow, as a result of Ward
identities that originate from charge conservation. It is very useful
to introduce the coupling constant $\gamma=\Gamma/Z$, which allows us
to rewrite the RG equations in a simpler way:
\begin{eqnarray}
\frac{dg}{d\ell} &=& \frac{1}{2}\,{g^2} - {g^2}
\left[2+2\;\frac{1-\gamma}{\gamma}\;\ln(1-\gamma)
\right]
\label{eqn:g-rg-2}\\
\frac{d\gamma}{d\ell} &=& g\,\gamma\;(1-\gamma)
\, .
\label{eqn:gamma-rg-2}
\end{eqnarray}
For $g>0$, it follows from Eq.~(\ref{eqn:g-rg-2}) that there are two
fixed-point values $\gamma^*=0,1$ (a closer analysis rules out the
possibility of another value of $\gamma^*$ with $g=0$), as shown in
Fig.~\ref{fig:flow}. The $\gamma^*=0$ fixed point is unstable, while
the $\gamma^*=1$ one is stable. Consider the RG equation for $g$. The
first term on the right-hand-side is the weak-localization correction,
while the second term is the interaction correction. The value $\gamma
= 0.42316\ldots$ separates the regime where the weak-localization
correction dominates over the interaction contribution ($dg/d\ell <0$
for $\gamma < 0.42316\ldots$ and $dg/d\ell >0$ for $\gamma >
0.42316\ldots$). Although the entire surface $g=0$ with arbitrary
$\gamma$ is left invariant under the RG flow, any system with bare
$g,\gamma\neq 0$ will necessarily flow into the $g=0,\gamma=1$ fixed
point. This is the case for short-range interactions, where the flow
starts with a value $\gamma<1$.  Note that if the bare interaction is
weak, $\gamma\ll 1$, then the resistivity will initially increase
before eventually decreasing to zero.

Now, consider the case of dynamically-screened Coulomb interactions.
As in the fermionic case, the Ward identity for charge conservation
requires the density-density correlation function to vanish
at ${\bf q}=0$. This, in turn, requires the ${\bf q}$-dependent
interaction $\Gamma(q)$, which generalizes $\Gamma$ to the case of
Coulomb interactions, to satisfy the
identity \cite{Finkelstein83,Castellani84,Belitz94}:
\begin{equation}
\label{eqn:Z-Gamma-identity}
Z - \Gamma(q) = \frac{\partial n}{\partial\mu}\,
\frac{q}{q + 4\pi {e^2} ({\partial n}/{\partial\mu})}
\end{equation}
Taking the $q\rightarrow 0$ limit of
(\ref{eqn:Z-Gamma-identity}), we obtain $Z=\Gamma$.
Substituting this identity into (\ref{eqn:g-rg}), we see that
the second term inside the square bracket
in (\ref{eqn:g-rg}) vanishes. Thus, the RG equation
for $g$ is $dg/d\ell=-3{g^2}/2$, and the resistivity flows
logarithmically to zero. The system is controlled
by the same infinite-conductivity fixed point
as in the short-ranged case.

Before concluding this section, let us write down the asymptotic behavior
near the fixed point $g^*=0,\gamma^*=1$, which we will need later to obtain
the critical exponents: $g\sim 2/3\ell$ and $1-\gamma \sim \exp(-\int
d\ell\; g)\sim \ell^{-2/3}$.

\section{Critical Behavior}

The most striking conclusion about the critical behavior of this system is
that the critical resistivity is zero!  In other words, the $2D$
superconductor-insulator transition is broadly similar to the $3D$ one.  This
is somewhat unexpected. In models such as the Bose-Hubbard model, which
describes a superfluid insulator transition in a clean system, or the
$2+1$-dimensional $XY$ model, one finds ${\sigma^*}=c\,{e^2}/h$, with $c$ a
finite universal number.  At our fixed point, $c=\infty$. Another odd feature
is the logarithmic approach to the critical resistivity which we find; this
logarithm is rather different from the type which are encountered in the
lower critical dimension of a phase transition (which happens to be $d=1$ for
the superfluid-insulator transition).  Since a logarithmic flow is rather
slow, it may not be possible to observe $c=\infty$. Instead, the critical
conductivity at a given temperature may actually appear to be a non-universal
number which depends on the bare conductivity.

Let us also consider the single-boson density of states, $N(\omega)$.
This may be studied by introducing a source term for $\text{tr}(\Lambda Q)$
into the effective action and computing its renormalization.
In a system with short-ranged interactions, we find:
\begin{equation}
\frac{d}{d\ell}\ln N = -\,g\cdot\gamma\cdot
\ln\!\left(1-\gamma\right)
\end{equation}
Substituting the asymptotic forms of $g$ and $\gamma$,
we find that the single-particle density of states diverges weakly,
$N(\omega)\sim e^{\frac{2}{9}\left[\ln\ln(1/\omega\tau)\right]^2}$.
Since the boson creation operator is the order
parameter for the superfluid phase and
\begin{equation}
N(\omega) = \text{Im}\,\overline{\left\langle {\psi^\dagger}({\bf x},\omega)\:
{\psi}({\bf x},-\omega)
\right\rangle}
\end{equation}
the scaling relation for $N(\omega)$ implies that the critical
exponent $\eta=0$ with logarithmic corrections.
However, in the presence of dynamically-screened Coulomb interactions,
there is a more severe divergence, and we find:
\begin{equation}
\frac{d}{d\ell}\ln N = \,g\cdot\ell
\end{equation}
Consequently, the single-boson density of states diverges
at the transition with the power-law
$N(\omega)\sim \omega^{-2/3}$. This implies that the critical
exponents $\eta$ and $z$ satisfy $\eta/z=-2/3$. Note that
we have calculated the density-of-states at a {\it metallic} critical
point. Thus, we should not expect Coulomb gap physics
to suppress it and give $\eta>0$. In the fermionic case,
the suppression of the density-of-states is due to the
dominance of the exchange interaction.

Our NL$\sigma$M does not explicitly include single-boson
operators. We assume that their properties can be deduced
from the the density-of-states. It is certainly possible
for single-particle operators to be critical even in a
theory in which only collective modes are retained;
this is the idea behind bosonization. It is conceivable,
however, that our NL$\sigma$M is incomplete, as regards
single-boson properties. This could occur if the critical
exponent controlling the correlation function
$\langle {\psi^\dagger}(x,0)\,{\psi}(x,0) \rangle$
were unrelated to that controlling
$\langle {\psi^\dagger}(0,\tau)\,{\psi}(0,0) \rangle$.

Since $Z$ diverges only logarithmically, the dynamical exponent, $z=2$, as
in a non-interacting system. However, in the case of
dynamically-screened Coulomb interactions, there are
actually two different diverging time scales. One, with
exponent $z$, is the scale associated with $Z$;
it controls the scaling of the specific heat and energy diffusion.
There is a second exponent, $z_c$, associated with $Z-\Gamma$,
which controls charge diffusion. By the same argument as in
a fermionic system \cite{Belitz94}, eq. \ref{eqn:Z-Gamma-identity}
implies at small $q$ that $Z-\Gamma \sim q$, from which
we conclude that $Z-\Gamma \sim \xi^{-1}$, i.e. ${z_c}=1$.
This result was obtained for
the superconductor-insulator transition by a closely-related
argument in ref. \onlinecite{Fisher89b}. Combining this
with our density-of-states calculation, we have $\eta=-2/3$
for Coulomb interactions. Notice that $\eta=-2/3<0$ satisfies the lower bound
$\eta<2-d$ of Ref. \onlinecite{Fisher89b} for $d=2$.
The density-of states and the dynamical exponent, $z_c$,
are the only quantities which distinguish short-ranged and
dynamically-screened Coulomb interactions in the infrared limit.

As we discussed in section IV, the leading perturbation of our $\sigma$ model
is a $\text{tr}\left(M\;Q\right)$ term, where $M$ is a constant matrix say in
replica space, which breaks the replica symmetry of the diffusive
saddle-point manifold possibly in the direction of the Bose-glass phase. This
is a dimension $2$ operator at tree level. (If the matrix $M$ is proportional
to the identity in replica space, this operator is instead just a constant at
the diffusive saddle-point.) Since the coupling constant $g$ flows to zero,
we expect a critical exponent $\nu=1/2$, up to logarithmic corrections.  This
value of $\nu$ -- the mean-field value -- violates the bound $\nu\geq 2/d$ of
ref. \onlinecite{Chayes86a}.  However, such violation has been seen in other
systems as well, and it has been argued \cite{Pazmandi97} that the exponent
bounded by the theorem of ref.  \onlinecite{Chayes86a} is, in fact, a
finite-size scaling exponent which can be different from $\nu$.

\section{Discussion}

Diffusion in two dimensions is marginal, and small corrections (in the limit
of large conductivity) such as that due to quantum interference or
interactions can tip the balance one way or the other. Contrary to
conventional wisdom, it is hardly a foregone conclusion which effect will
win. After all, weak localization is {\it weak}.  Interactions can easily
overpower it, leading to metallic behavior.  According to our analysis, this
is precisely what occurs at the superconductor-insulator transition.  The
effect of interactions is so dominant that the universal value of the
conductivity at the transition is infinity. Such a diverging conductivity has
been found in models with interaction and dissipation, but without
disorder~\cite{Phillips1}.

The possibility of a metallic phase within the Bose glass phase has been
studied recently \cite{Doniach,Phillips2}. We focus on the diffusive
properties at the critical point, and do not investigate whether saddle-point
solutions whithin the Bose glass could lead to non-zero conductivities.
However, it is noteworthy that an infinite critical conductivity is
consistent with a Bose metal with a diverging conductivity at the transition
\cite{Phillips2}.

We derive these results in a NL$\sigma$M approach, in
which we discard those critical modes of
the clean system which are extraneous
and retain only the particle diffusion modes of the
disordered system. The resulting NL$\sigma$M
leads to a number of non-trivial predictions:
(1) the critical conductivity is infinite; (2)
there are two diverging times scales if the interaction
is Coulombic, one associated with
charge diffusion, which has exponent $z=1$,
the other associated with energy diffusion, which has exponent $z=2$;
(3) the single-boson density of states diverges as $\omega^{-2/3}$,
which implies a critical exponent $\eta=-2/3$ in the case
of Coulomb interactions; for short-range interactions, it
diverges logarthmically; (4) the correlation-length exponent
takes the mean-field value $\nu=1/2$.

If boson-vortex duality were to hold exactly, then one would expect ${g^*}=1$
(in units of ${(2e)^2}/h$).  Our result appears to imply that duality is
violated logarithmically: bosons are more mobile than vortices in the
infrared limit. However, it is hard to see how the physics of vortices enters
at all into our calculation, so it is possible that we have missed important
non-perturbative effects. Our results do not agree with the numerical study
of Wallin, {\it et al.} \cite{Wallin94}.  However, the flow to our fixed
point is logarithmic, and this may be too slow for a numerical study on a
finite-sized system.  Alternatively, they may simply be accessing a different
fixed point which attracts systems with small bare conductivities. And
finally, since their starting point studies phase but no amplitude
fluctuations, the two models may simply be in different universality classes.
Our results also differ quantitatively from those of Herbut,
which are based on an expansion about $d=1$.\cite{Herbut00}

The measured critical exponents for the zero-field
superconductor-insulator transition, which is
accessed by varying the thickness of a thin film
\cite{Haviland89,Markovic99}, are those of classical
percolation. This does not agree with our theory,
but it also suggests that the experiments are not
quite in the asymptotic quantum critical regime,
but rather in some higher-temperature classical regime.
There is disagreement about the values of
the critical exponents at the 
magnetic-field-tuned superconductor-insulator transition.
One experiment \cite{Hebard90} finds percolation-like exponents,
while another \cite{Markovic99} finds 
$\nu=0.7\pm 0.2$, which includes our theoretical
prediction at the edge of its error bar.
(All of these experiments find $z\approx 1$, as expected
on general grounds \cite{Fisher89b,Belitz94}, and in our theory.)
The applicability of our strategy to a magnetic-field-tuned
superconductor-insulator transition is a
question for future study.

\begin{acknowledgments}
We would like to thank M.P.A. Fisher, V. Gurarie,
A.W.W. Ludwig, and P. Phillips for discussions.
C.N. was supported by the National Science Foundation under
Grant No. DMR-9983544 and by the A.P. Sloan Foundation. C.C.
was supported by the National Science Foundation under
Grant No. DMR-9876208 and by the A.P. Sloan Foundation.
\end{acknowledgments}

\appendix

\begin{widetext}
\section{Derivation of the $\sigma$-model}
\label{appendix-Q-sigma}

Here we derive the effective action for the interacting disordered bosons in
terms of two fields $\Phi$ and $Q$. We work, in sequence, on the free part,
the disorder part, and finally the interaction part of Eq.
(\ref{eq:boson-action}).

\subsection{The free action}
We start by introducing a bosonic amplitude $\Phi$ to decouple the chemical
potential ($\mu$) term. $\Phi$ acquires a finite expectation value when
bosons condense. 

The free part of the action
\begin{equation}
S_{\text {free}}[\phi]= {\sum_{n,m}}\int {d^2}x\;i\;{\phi_{naA}}
(x)\left(i{\epsilon_n}+
\frac{1}{2m}{\nabla^2}+\mu\right)\,
{\Lambda_{nm}}\;{\phi_{maA}}(x)
\end{equation}
can is generated upon integration of a decoupling field
$\Phi_{na}=\Phi_{na1}+i\Phi_{na2}$ in
\begin{eqnarray}
S_{\text {free}}[\phi,\Phi]&&= {\sum_{n,m}}\int {d^2}x\;i\;{\phi_{naA}}
(x)\left(i{\epsilon_n}+
\frac{1}{2m}{\nabla^2}\right)\,
{\Lambda_{nm}}\;{\phi_{maA}}(x) 
\nonumber\\
&&+{\sum_{n}}\int {d^2}x \; \frac{1}{2}\Phi_{na}^*\Phi_{na}
+{\sum_{n}}\int {d^2}x \;\sqrt{2\mu}\;{\phi_{naA}}\;
\left[{e^{-i{\frac{\pi}{4} \text{sgn}({n})}}}\right]\;\Phi_{naA}
\label{eq:free}
\end{eqnarray}

\subsection{Disorder term}
Let us next decouple the four bosons in the disorder term in Eq.
(\ref{eq:boson-action}):
\begin{equation}
S_{\text {rand}}= {\sum_{n,n',m,m'}}
\int {d^2}x\: \frac{v_0}{2}\: {\phi_{naA}}(x){\Lambda_{nn'}}{\phi_{n'aA}}(x)
\;{\phi_{mbB}}(x){\Lambda_{mm'}}{\phi_{m'bB}}(x)
\end{equation}
where ${\Lambda_{mm'}}=\text{sgn}(m)\,\delta_{mm'}$.
%
%
The same disorder term is generated upon integration of the
Hubbard-Stratonovich matrix field $Q^{mn}_{ab,AB}$
\begin{equation}
e^{-S_{\rm rand}[\phi]}=
\int DQ \;\;
e^{-\frac{1}{2v_0}
\int d^d x \;{\rm tr}\; Q^2}\;
e^{-S_{\rm HS}[Q,\phi]}
\end{equation}
where
\begin{equation}
S_{\rm HS}[Q,\phi]= i{\sum_{n,m,m'}}
\int {d^2}x\: 
\: {\phi_{naA}}(x)
\; Q^{nm}_{ab,AB}(x)\;{\Lambda_{mm'}}
{\phi_{m'bB}}(x)
\end{equation}
The matrix $Q$ has indices in three separate spaces, {\it i.e.}, it is
assembled as a direct product in energy $n,m$, replica $a,b$, and
real-imaginary $A,B$ spaces. The trace of $Q^2$ corresponds to
\begin{equation}
{\rm tr} \; Q^2=Q^{nm}_{ab,AB}\;Q^{mm}_{ba,BA}\; ,
\end{equation}
where repeated index summation is carried out in all three spaces.  When we
write for short $Q_{nn'}$ we mean a matrix whose elements are matrices in
replica, and real-imaginary spaces. 

%
%
\subsection{Interaction term}
Let us consider the case of short range interactions, in the
density-density (s) and pairing (c) channel. Once again, we will omit
sums over indices for replica and real-imaginary parts, and write
explicitly the Matsubara sums.
\begin{equation}
S_{\rm int}=S_{s}+S_{c}
\end{equation}
where
\begin{eqnarray}
S_{s}&=&{\Gamma_s}
\sum_{n_1,\dots,n_4} \int d^d x 
\: {e^{-i\frac{\pi}{4} \sum_j\text{sgn}({n_j})}}
\left[
{\phi_{n_1aA}}(x)
\;J_{AA'}\;
{\phi_{n_2aA}}(x)
\right]
\left[
{\phi_{n_3aB}}(x)
\;J_{BB'}\;
{\phi_{n_4aB}}(x)
\right]
\;
\delta_{n_1-n_2+n_3-n_4}
\\
S_{c}&=&{\Gamma_c}
\sum_{n_1,\dots,n_4} \int d^d x 
\; {e^{-i\frac{\pi}{4} \sum_j\text{sgn}({n_j})}}
\left[
{\phi_{n_1aA}}(x)
\;S_{AA'}^+\;
{\phi_{n_2aA'}}(x)
\right]
\left[
{\phi_{n_3aB}}(x)
\;S_{BB'}^-\;
{\phi_{n_4aB'}}(x)
\right]
\;
\delta_{n_1+n_2-n_3-n_4}
\end{eqnarray}
with the matrices $J_{AB}=\frac{1}{\sqrt{2}}(\delta_{AB} -\sigma_{AB}^2)$ and 
$S_{AB}^\pm=\sigma_{AB}^3 \pm i\sigma_{AB}^1$ (the
$\sigma^{i}$ being Pauli matrices). Notice that the different terms
within square brackets above correspond, in terms of the original
bosons $\psi$, to $\psi^*\psi$, $\psi^*\psi^*$, and $\psi\psi$.

We now introduce two Hubbard-Stratonovich fields $X$ and ${X_c},{X_c}^*$ to
decouple the four $\phi$ interactions:
\begin{eqnarray}
e^{-S_s[\phi]}&=&
\int DX \;
e^{-S_{\rm y}[X]}\;
e^{-S_{\rm HS/X}[X,\phi]}\\
S_{\rm HS/X}[X]&=&
i\sqrt{2\Gamma_s}\;
\sum_{n,m}
\;\int d^d x
\;
{\phi_{naA}}(x)
\; {e^{-i\frac{\pi}{4} \text{sgn}({n})}}
\; X^{nm}_{ab,AB}(x)\;
\; {e^{-i\frac{\pi}{4} \text{sgn}({m})}}
{\phi_{mbB}}(x)
\label{eq:Sy}
\end{eqnarray}
and
\begin{eqnarray}
e^{-S_c[\phi]}&=&
\int D{X_c}^*D{X_c} \;
e^{-S_{\rm z}[{X_c}]}\;
e^{-S_{\rm HS/{X_c}}[{X_c},\phi]}\\
S_{\rm HS/{X_c}}[{X_c}]&=&
i\sqrt{2\Gamma_c}\;
\sum_{n,m}
\;\int d^d x
\;
{\phi_{naA}}(x)
\; {e^{-i\frac{\pi}{4} \text{sgn}({n})}}
\;\frac{1}{2}\left[
{X_c}^{nm}_{ab,AB}(x) + {{X_c}^\dagger}^{nm}_{ab,AB}(x)
\right]
\; {e^{-i\frac{\pi}{4} \text{sgn}({m})}}
{\phi_{mbB}}(x)
\label{eq:Sx}
\end{eqnarray}
where
\[
X^{nm}_{{ab},AB}=
{\rm X}^{n-m}_{a}
\;\delta_{ab}\;J_{AB}
\qquad
{X_c}^{nm}_{{ab},AB}=
{\rm {X_c}}^{n+m}_{a}
\;\delta_{ab}\;S_{AB}^+
\]
Notice that the matrices $X^{nm}_{{ab},AB}$ and ${X_c}^{nm}_{{ab},AB}$
depend, respectively, only on the energy difference $n-m$ and sum
$n+m$. The action for the matrices $X$ and ${X_c}$ is
\begin{eqnarray}
S_{\rm x}[X]&=&\frac{1}{2}
\sum_{n}
\int d^d x
\;
\; {\rm X}^{n}_{a}
\;\; {\rm X}^{-n}_{a}
\label{eq:X}
\\
S_{{\rm x}_c}[{X_c}]&=&\frac{1}{2}
\sum_{n}
\int d^d x
\;
\; {\rm {X_c}^*}^{n}_{a}
\; {\rm {X_c}}^{n}_{a}
\label{eq:X_c}
\end{eqnarray}

\subsection{Integrating out the $\phi$ fields}

We can summarize all terms above:
\begin{eqnarray}
{X_c}=&&\int D\phi \;D\Phi DQ DX D{X_c}^*D{X_c} \; 
e^{-\frac{1}{2v_0}
\int d^d x \;{\rm tr}\; Q^2}\;
e^{-S_{\rm x}[X]}\;
e^{-S_{{\rm x}_c}[{X_c}]}\;
\nonumber \\
&&\,\,\,\,\, \times
e^{-S_{\rm free}[\phi,\Phi]}\;
e^{-S_{\rm HS}[\phi,Q]}\;
e^{-S_{\rm HS/X}[\phi,X]}\;
e^{-S_{\rm HS/{X_c}}[\phi,{X_c}]}\;
\label{eq:sigma-int}
\ .
\end{eqnarray}
where we can express the $S_{\rm free},S_{\rm HS},S_{\rm HS/X}$, and
$S_{\rm HS/{X_c}}$ in a more concise (matrix) notation as follows:
\begin{eqnarray}
S_{\rm free}[\phi,\Phi]&=&
\int {d^2}x\;i\;{\phi}^T(x)\left(i{\Omega}+
\frac{1}{2m}{\nabla^2}\right)\,
{\Lambda}\;{\phi}(x) 
+
\sqrt{2\mu}\; \int {d^2}x \; \phi^T\;{e^{-i\frac{\pi}{4} \Lambda}}
\;\Phi
\\
S_{\rm HS}[\phi,Q]&=&
\int {d^2}x\: 
\: {\phi}^T(x)
\; i Q(x)\;{\Lambda}
{\phi}(x)
\\
S_{\rm HS/X}[X]&=&
i\sqrt{2\Gamma_s}\;
\;\int d^d x
\;
{\phi}^T(x)
\;{e^{-i\frac{\pi}{4}\Lambda}}
\; X(x)\;
\;{e^{-i\frac{\pi}{4}\Lambda}}
{\phi}(x)
\\
S_{\rm HS/{X_c}}[{X_c}]&=&
i\sqrt{2\Gamma_c}\;
\;\int d^d x
\;
{\phi}^T(x)
\;{e^{-i\frac{\pi}{4}\Lambda}}
\; \frac{1}{2}\left[ {X_c}(x)+{X_c}^\dagger(x) \right]\;
\;{e^{-i\frac{\pi}{4}\Lambda}} 
{\phi(x)}
\end{eqnarray}
where the matrix $\Omega_{nm}=\epsilon_n\;\delta_{nm}$.

Integrating out the boson fields $\psi$, we obtain
\begin{equation}
{X_c}=\int D\phi \;D\Phi DQ DX D{X_c}^*D{X_c} \; 
e^{-\frac{1}{2v_0}
\int d^d x \;{\rm tr}\; Q^2}\;
e^{-S_{\rm x}[X]}\;
e^{-S_{{\rm z}_c}[{X_c}]}\;
\;e^{-S_0[Q,\Phi,X,{X_c}]}
\end{equation}
with
\begin{eqnarray}
S_0[Q,\Phi,X,{X_c}]=
&&
\mu \int d^d x\;
\Phi^T\;G\;\Phi
+\int d^d x\;
\;{\text {tr}}\log
\Big[
\;i\left(i{\Omega}+
\frac{1}{2m}{\nabla^2}\right)\,
{\Lambda}\;
+
iQ(x)\;\Lambda
\nonumber\\
&&+
i\sqrt{2\Gamma_s}
\;{e^{-i\frac{\pi}{4}\Lambda}}
\; X(x)\;
\;{e^{-i\frac{\pi}{4}\Lambda}}
+
i\sqrt{2\Gamma_c}
\;{e^{-i\frac{\pi}{4}\Lambda}}
\; \frac{1}{2} \left( {X_c}(x)+{X_c}^\dagger(x) \right)\;
\;{e^{-i\frac{\pi}{4}\Lambda}}
\Big]
\\
=&&
\mu \int d^d x\;
\Phi^T\;G\;\Phi
+\int d^d x\;
\;{\text {tr}}\log
\Big[
\left(i{\Omega}+
\frac{1}{2m}{\nabla^2}\right)\,
+
Q(x)
\nonumber\\
&&
-i\sqrt{2\Gamma_s}
\;{e^{-i\frac{\pi}{4}\Lambda}}
\; X(x)\;
\;{e^{+i\frac{\pi}{4}\Lambda}}
-i\sqrt{2\Gamma_c}
\;{e^{-i\frac{\pi}{4}\Lambda}}
\; \frac{1}{2} \left( {X_c}(x)+{X_c}^\dagger(x) \right)\;
\;{e^{+i\frac{\pi}{4}\Lambda}}
\Big] + {\rm const.}
\end{eqnarray}
The propagator $G$ depends on $Q,X$, and ${X_c},{X_c}^*$.

\subsection{Shifting $Q$}

Let us now shift $Q$:
\[
Q \to {\tilde Q}=
Q
-i\sqrt{2\Gamma_s}
\;{e^{-i\frac{\pi}{4}\Lambda}}
\; X\;
\;{e^{+i\frac{\pi}{4}\Lambda}}
-i\sqrt{2\Gamma_c}
\;{e^{-i\frac{\pi}{4}\Lambda}}
\; \frac{1}{2} \left( {X_c}+{X_c}^\dagger \right)\;
\;{e^{+i\frac{\pi}{4}\Lambda}}
\]
and
\begin{eqnarray}
{\rm tr} Q^2
=
&\;&
{\rm tr}\; {\tilde Q}^2 \nonumber\\
&\;& 
+i2\sqrt{2\Gamma_s}\; {\rm tr}\;\left[
{\tilde Q} \;{e^{-i\frac{\pi}{4}\Lambda}}
\; X\;
\;{e^{+i\frac{\pi}{4}\Lambda}}\right]
+
i2\sqrt{2\Gamma_c}\; {\rm tr}\;\left[
{\tilde Q} \;{e^{-i\frac{\pi}{4}\Lambda}}
\; \frac{1}{2} ({X_c}+{X_c}^\dagger)\;
\;{e^{+i\frac{\pi}{4}\Lambda}}\right]
\\
&\;&
-{4\Gamma_s}\;2k\sum_{n}
\int d^d x
\;
\; {\rm X}^{n}_{a}
\;\; {\rm X}^{-n}_{a}
-{4\Gamma_c}\;2k\sum_{n}
\int d^d x
\;
\; {\rm {X_c}^*}^{n}_{a}
\; {\rm {X_c}}^{n}_{a}
\, .
\end{eqnarray}
The Matsubara cut-off $k$ comes from the extra frequency sum in the
trace, and there are factors of 2 from the traces over the
real-imaginary components, $\text{tr}\; J=2$ and $\text{tr} \;S^+S^-=4$.

The next step is to integrate out the $X,{X_c}$ fields. This generates quadratic

in $Q$ terms (we now drop the tildes for notational simplicity). It is useful
to define
\[
\tilde\Gamma_{s,c}=
\frac{\Gamma_{s,c}}{v_0^2}\;\frac{1}{1-8\;\frac{\Gamma_{s,c}k}{v_0}}\ .
\]
\begin{equation}
S_{\rm fink}=
\tilde\Gamma_{s,c}
\sum_{n_1,\dots,n_4}\;
\int d^d x\;
\left[
{e^{+i\frac{\pi}{4}\text{sgn}\;n_1}}\;
Q^{n_1n_2}_{aa,AA'} \;{e^{-i\frac{\pi}{4}\text{sgn}\;n_2}}
\right]\;
\gamma^{s,c}_{AA',BB'}\;
\left[{e^{+i\frac{\pi}{4}\text{sgn}\;n_3}}\;
Q^{n_3n_4}_{aa,BB'} \;{e^{-i\frac{\pi}{4}\text{sgn}\;n_4}}
\right]\;
\delta_{n_1\mp n_2\pm n_3+n_4}
\end{equation}
where the tensors $\gamma_{AA',BB'}$ depend on the channel:
\begin{eqnarray}
\gamma^s_{AA',BB'}
&=&
J_{AA'}\;J_{BB'}\;
\\
\gamma^c_{AA',BB'}
&=&
S^+_{AA'}\;S^-_{BB'}\;
\end{eqnarray}

Summarizing it all, we have an effective action
\begin{eqnarray}
&&S_{\text eff}[Q,\Phi]=
\frac{1}{2v_0}
\int d^d x \;{\rm tr}\; Q^2
+\int d^d x\;
\;{\text {tr}}\log
\Big[
\left(i{\Omega}+
\frac{1}{2m}{\nabla^2}\right)\,
+
Q(x)\Big] +\mu \int d^d x\;
\Phi^T\;G\;\Phi
\\
&&\;\;\;\;\;\;\;\;\;
+
\tilde\Gamma_{s}
\sum_{n_1,\dots,n_4}\;
\int d^d x\;
\left[
{e^{+i\frac{\pi}{4}\text{sgn}\;n_1}}\;
Q^{n_1n_2}_{aa,AA'} \;{e^{-i\frac{\pi}{4}\text{sgn}\;n_2}}
\right]\;
\gamma^{s}_{AA',BB'}\;
\left[{e^{+i\frac{\pi}{4}\text{sgn}\;n_3}}\;
Q^{n_3n_4}_{aa,BB'} \;{e^{-i\frac{\pi}{4}\text{sgn}\;n_4}}
\right]\;
\delta_{n_1-n_2+n_3-n_4}
\nonumber\\
&&\;\;\;\;\;\;\;\;\;
+
\tilde\Gamma_{c}
\sum_{n_1,\dots,n_4}\;
\int d^d x\;
\left[
{e^{+i\frac{\pi}{4}\text{sgn}\;n_1}}\;
Q^{n_1n_2}_{aa,AA'} \;{e^{-i\frac{\pi}{4}\text{sgn}\;n_2}}
\right]\;
\gamma^{c}_{AA',BB'}\;
\left[{e^{+i\frac{\pi}{4}\text{sgn}\;n_3}}\;
Q^{n_3n_4}_{aa,BB'} \;{e^{-i\frac{\pi}{4}\text{sgn}\;n_4}}
\right]\;
\delta_{n_1+n_2-n_3-n_4}
\nonumber
\end{eqnarray}

\end{widetext}
\section{Separation of longitudinal and transverse (diffusive) modes}
\label{appendix-diffusion}

In this appendix we show that the transverse fluctuations of the $Q$
field correspond to boson diffusion for $\mu=0$, similarly to the
fermionic case.  We show that the diffusion term arises even in the
absence of a small parameter $1/E_F\tau$.

Expansion to quadratic order in $\delta Q$ leads to a term
\begin{equation}
\frac{1}{2}\;M^{n_1n_2}(q)
\;\;\delta Q^{n_1n_2}_{aa',AA'}(-q)  \;\delta Q^{n_2n_1}_{a'a,A'A}(q)
\end{equation}
where
\begin{equation}
M^{n_1n_2}(q)
=
\frac{1}{v_0}-\int \frac{d^dp}{(2\pi)^d}\;G(p,n_1)\;G(p+q,n_2)
\; .
\label{eq:M}
\end{equation}
The first term on the right-hand side comes from the ${\text tr} Q^2$ in the
action, and the second term has its origin in the ${\text tr}\log(\cdot)$. 
$M^{n_1n_2}(q)$ is selected to be diagonal in and independent of replica
$a,a'$ and real-imaginary $A,A'$ indices. The Green's function
\begin{equation}
G(p,n_1)=\frac{1}{i\epsilon_n -E(p)+\frac{i}{2\tau}\;\text{sgn}(\epsilon_n )}
\; .
\end{equation}
For $\text{sgn}n_1 \;\text{sgn}n_2>0$, the real part of the integral vanishes
for $q\to 0$, so that $\text{Re}[\;M^{n_1n_2}(q)] = \frac{1}{v_0}$, and we are
left with a massive longitudinal mode.

Let us turn to the interesting case $\text{sgn}n_1\;\text{sgn}n_2<0$. For
simplicity, we neglect the $i\epsilon_{n_{1,2}}$ terms in the denominator
(these terms can be handled alternatively by shifting the $Q$ field).
Expanding the integral in Eq.~(\ref{eq:M}) in powers of $q$:
\begin{multline}
\int\frac{d^dp}{(2\pi)^d}\;G^0(p,n_1)\;G^0(p+q,n_2)=\\
\!\!\!\!\!\!\!\!\!\!\!\!\!\!\!\!
\!\!\!\!\!\!\!\!\!\!\!\!\!\!\!\!
\!\!\!\!\!\!\!\!\!\!\!\!\!\!\!\!
\int\frac{d^dp}{(2\pi)^d}\;G_+(p)\;G_-(p)
\\
+
\frac{1}{2}\frac{q^2}{m}\;
\int\frac{d^dp}{(2\pi)^d}
\bigg[
G_+^2(p)\;G_-(p)
\\
+
\frac{4E(p)}{d}\;G_+^3(p)\;G_-(p)
\bigg]
\end{multline}
where $G_\pm(p)=[-E(p)\pm \frac{i}{2\tau}]^{-1}$. The integrals over
momenta can be transformed into integrals over energy $\epsilon$ using the
density of states
$\nu(\epsilon)=\frac{1}{2}\frac{S_d}{(2\pi)^d}\;(2m)^{d/2}\epsilon^{d/2-1}$.
Define
\begin{equation}
I_{a,b,c}=\int_0^\Omega d\epsilon\;\nu(\epsilon)\;
\epsilon^a\;[G_+(\epsilon)]^b\;[G_-(\epsilon)]^c
\; ,
\end{equation}
so 
\begin{multline}
\int\frac{d^dp}{(2\pi)^d}\;G^0(p,n_1)\;G^0(p+q,n_2)=\\
I_{0,1,1}+\frac{1}{2}\frac{q^2}{m}\;
\bigg[I_{0,2,1}+\frac{4}{d}I_{1,3,1}\bigg]
\; .
\end{multline}
(a finite upper frequency cut-off $\Omega$ is needed depending on
$d,a,b,c$).  It is also convenient to rescale the energies, defining
$y=2\tau\epsilon$, so we can write
\begin{multline}
I_{a,b,c}=\frac{1}{2}\frac{S_d}{(2\pi)^d}\;(2m)^{d/2}\;
({2\tau})^{-d/2-a+b+c}\\
\times
\int_0^{2\tau\Omega} dy\;
\frac{y^{d/2-1+a}}{(-y+i)^{b}\;(-y-i)^{c}}
\\
={\cal A}_d\;({2\tau})^{-a+b+c}\;
\int_0^{2\tau\Omega} dy\;
\frac{y^{d/2-1+a}}{(-y+i)^{b}\;(-y-i)^{c}}
\; ,
\end{multline}
with ${\cal A}_d=\frac{1}{2}\frac{S_d}{(2\pi)^d}\;(2m)^{d/2}\;
({2\tau})^{-d/2}$.

One can check that once $\tau$ is fixed by the saddle point
Eq.(\ref{eqn:S-P-condition}), which can be cast as
\begin{equation}
\text{Im}\; I_{0,1,0}=-\text{Im}\; I_{0,0,1}=\frac{1}{2\tau v_0}
\; ,
\end{equation}
then it follows trivially that
\begin{equation}
I_{0,1,1}=\frac{1}{v_0}
\; ,
\end{equation}
so that the leading order term in $M_\perp(q)$ is of order $q^2$, which
allows us to define the diffusion constant
\begin{equation}
D=\frac{1}{4m}\bigg[I_{0,2,1}+\frac{4}{d}I_{1,3,1}\bigg]
\; .
\end{equation}
The last step remaining is to show that $D$ is {\it purely real}. After
simple manipulations, one can show that
\begin{multline}
\text{Im}D=
\frac{1}{4m}\frac{S_d}{(2\pi)^d}\;(2m)^{d/2}\;
({2\tau})^{-d/2+3}\\
\times
\int_0^{2\tau\Omega} dy\;\left[\left(\frac{8}{d}-1\right)y^2-1 \right]\;
\frac{y^{d/2-1}}{(y^2+1)^3}
\; .
\label{eq:imD}
\end{multline}
It is trivial to show by integration by parts (splitting the
integrands into $f(y)=y/(y^2+1)^3$ and $g(y)=y^\alpha$) that the
integral in Eq.~(\ref{eq:imD}) scales as $(\tau\Omega)^{d/2-4}$.  Thus
the cut-off can be safely taken to infinite for $d<8$, and 
$\text{Im}D=0$.

Notice the difference between the fermionic and bosonic cases. In the
fermionic case one can also interchange momentum $p$ integrals for
energy $\epsilon$ integrals, using the density of states at the Fermi
level $E_F$. The integrals are cut-off by the bottom of the band,
$-E_F$ away from the zero energy states. In the bosonic case, one
starts from the bottom of the band, and needs to include an energy
dependent density of states $\nu(\epsilon)$; the cut-off $\Omega$ is
introduced only for convergence, and $\Omega\to\infty$ is possible for
$d<8$. In contrast to the fermionic case, where $E_F$ is finite, in
the bosonic case for a perfect parabolic spectrum $\Omega\to\infty$.
The small parameter for the Fermi case is $(E_F\tau)^{-1}$, whereas for
the Bose case it is $(\Omega\tau)^{-1} \to 0$.

\section{Parameterization of the saddle and relation to the 
fermionic $\sigma$-model}
\label{appendix-saddle-param}

As we previously mentioned, we can easily obtain the RG equations for
the conductance and interaction couplings by determining a
correspondence with the fermionic model. Here we show how this is
achieved.

Let us first look at the Finkelstein type terms in the effective
action for the $Q$ fields. The $Q$ matrices are parameterized as in in
Eq.(\ref{eqn:Q-param}), repeated here for convenience.
\begin{eqnarray}
{Q} = \frac{m v_0}{2}\,
\left(\begin{array}{lccr}
i\left(1+q{q^T}\right)^{\frac{1}{2}} & q \\
{q^T} &  -i\left(1+{q^T}q\right)^{\frac{1}{2}}\\
                      \end{array} \right)
\end{eqnarray}
The quantities that appear in the Finkelstein type terms for the
bosonic problem are
\begin{eqnarray}
&&{e^{+i\frac{\pi}{4}\Lambda}}\;
Q\;{e^{-i\frac{\pi}{4}\Lambda}}
= \nonumber\\
&&
\frac{m v_0}{2}\,
\left(\begin{array}{lccr}
i\left(1+q{q^T}\right)^{\frac{1}{2}} & iq \\
-i({q^T}) &  -i\left(1+{q^T}q\right)^{\frac{1}{2}}\\
                      \end{array} \right)\\
&&
\frac{m v_0}{2}\,
i\left(\begin{array}{lccr}
\left(1-q(-{q^T})\right)^{\frac{1}{2}} & q \\
(-{q^T}) &  -\left(1-(-{q^T})q\right)^{\frac{1}{2}}\\
                      \end{array} \right)
\label{eq:sign-int}
\end{eqnarray}

Direct comparison with the fermionic saddle point
\begin{eqnarray}
{Q_F} = \frac{m v_0}{2}\,
\left(\begin{array}{lccr}
\left(1-q{q^T}\right)^{\frac{1}{2}} & q \\
{q^T} &  -\left(1-{q^T}q\right)^{\frac{1}{2}}\\
                      \end{array} \right)
\end{eqnarray}
shows that the terms in the Finkelstein type action for bosons, upon
parameterization in terms of $q,q^T$, are the same as the ones for fermions
upon the identification $q\to q$ and $q^T\to -q^T$. The extra factor of $i$
in \ref{eq:sign-int}, once squared (because the Finkelstein terms are
quadratic in $Q$), makes the sign of the interaction term for bosons and
fermions the same.

For discussing the diffusive term $D\int d^d x {\rm tr} (\nabla Q)^2$,
notice that by rewriting
\begin{eqnarray}
{Q} = \frac{m v_0}{2}\,
i \left(\begin{array}{lccr}
\left(1-q(-{q^T})\right)^{\frac{1}{2}} & -iq \\
i(-{q^T}) &  -\left(1-(-{q^T})q\right)^{\frac{1}{2}}\\
                      \end{array} \right)
\end{eqnarray}
we again identify it with the fermionic saddle point $Q_F$, but now
the off-diagonal elements have extra factors $i,-i$. These factors
will cancel each other in the expansion of the quadratic in $Q$
diffusive term, and hence can be dropped, and once again the fermionic
saddle expansion can be used. The overall factor of $i$ has the effect
of changing $D\to -D$.

In summary, all the RG equations for the dirty interacting boson
problem can obtained from those of the (interacting) fermionic
orthogonal ensemble upon replacing $g\to -g$ (or $D\to -D$).

%
%
%


\end{document}